# Real-time phase-retrieval and wavefront sensing enabled by an artificial neural network

JONATHON WHITE,[1,*] SICI WANG,[1,2] WILHELM ESCHEN,[1,2], AND JAN ROTHHARDT[1,2,3]

[1]*Institute of Applied Physics, Abbe Center of Photonics, Friedrich-Schiller-University Jena, Albert-Einstein-Straße 15, 07745 Jena, Germany*
[2]*Helmholtz-Institute Jena, Fröbelstieg 3, 07743 Jena, Germany*
[3]*Fraunhofer Institute for Applied Optics and Precision Engineering, Albert-Einstein-Str. 7, 07745 Jena, Germany*
[*]*jonathonwhite@protonmail.com*

**Abstract**

In this manuscript we demonstrate a method to reconstruct the wavefront of focused beams from a measured diffraction pattern behind a diffracting mask in real-time. The phase problem is solved by means of a neural network, which is trained with simulated data and verified with experimental data. The neural network allows live reconstructions within a few milliseconds, which previously with iterative phase retrieval took several seconds, thus allowing the adjustment of complex systems and correction by adaptive optics in real time. The neural network additionally outperforms iterative phase retrieval with high noise diffraction patterns.

## 1. Introduction

Artificial neural networks have become an effective solution for solving many problems. For example, neural networks are used for pattern recognition [1] (e.g. face identification) or medical diagnosis [2, 3]. Furtheremore, artificial neural networks have become an effective solution for solving many problems in optics, particularly phase retrieval [4–7]. This is mainly due to increasingly affordable graphics cards computational power required to train neural networks, and the recent advancements in neural network architecture design and training. CDI (Coherent Diffractive Imaging) is a lensless imaging method used to image an object by measuring the diffraction pattern which it produces when illuminated with coherent radiation source. Up to now iterative phase retrieval algorithms are employed for the necessary phase retrieval on the measured data, which are time consuming. The ability for neural networks to out-perfrom iterative CDI significantly in terms of retrieval speed has been proven [8]. In this contribution we apply a neural network to the field of CDI.

Recently, Neural networks have been used in application to CDI [6] and ptyography [7] as a new alternative to the iterative method for CDI [9] and therefore allow significantly faster reconstruction of the phase and intensity of objects from measured diffraction patterns.
In this work, we present the first application of a neural network to the phase retrieval of measured experimental data of a CDI-based wavefront sensor [10]. Our demonstration paves the way for real-time lensless wavefront sensing, particularly at short XUV and X-ray wavelengths, where lensless imaging techniques provide nanoscale resolution beyond the capabilities of imaging optics. Access to a real-time phase retrieval may help alignment of complex optical systems with feedback in real-time.

## 2. Basics of Lenseless Imaging and Phase Retrieval

Lenseless coherent diffractive imaging is used to characterize objects and coherent beams based on their diffraction pattern (Fig. 1 c). The advantage of this method is this is a form of imaging with no lenses required, and therefore the quality of imaging is not dependent on the quality of optics. This is particularly important in the extreme ultraviolet and X-ray spectral region, where high-quality optics are difficult to fabricate and lensless microscopes have readily demonstrated few-nm resolution beyond the capabilities of today's optics [11, 12]. Since only an intensity image is recorded, a phase retrieval method needs to be employed in the image reconstruction. In conventional coherent diffractive imaging (CDI) the constraint is an area around the object that is known to have zero transmission (support constraint) [9]. A known transmission function of the object enables the characterization of the input beam in amplitude and phase, which we term wavefront sensing. Recently, wavefront sensing with high accuracy has been demonstrated in the XUV wavelength [10]. To perform retrieval of the objects exit wave with the iterative method, a constraint must first be applied to the object for the algorithm to converge. For our wavefront sensing application the object is a transmission mask consisting of a thin, absorbing metal film on a silicon nitride membrane with holes. This mask is placed in the path of the laser beam, the laser beam passes through the holes and a diffraction pattern is produced which is imaged in the far field by a CCD camera.

The transmission function is 1 in the holes and 0 outside, thus the object exit wave is isolated and the structure can be directly applied a support constraint for iterative phase retrieval. By using a phase retrieval to find the object, which produced the measured diffraction pattern, we characterize the input beam. The iterative phase retrieval algorithm (Fig. 1) for CDI uses an initial guess for the object, and applies a support constraint in the spatial (object) domain as well as a Fourier constraint in the frequency domain. The support constraint in the object domain forces the retrieved object to fit within the support area, and the Fourier constraint in the frequency domain forces the retrieved object diffraction pattern to match the measured diffraction pattern. These constraints are applied iteratively, and each iteration involves two Fourier transforms between object and frequency domain. To measure the object behind the wavefront sensor using the iterative retrieval CDI method, an interpolation must be performed because the masked object (Fig. 2 d) will be retrieved. In contrast to an iterative algorithm, the neural network is able to retrieve the object from a diffraction pattern using a learned spatial support constraint without any iterative algorithm. It performs a retrieval by a singlepass through the neural network, which is a series of matrix multiplications and therefore is able to retrieve the phase in the order of milliseconds.

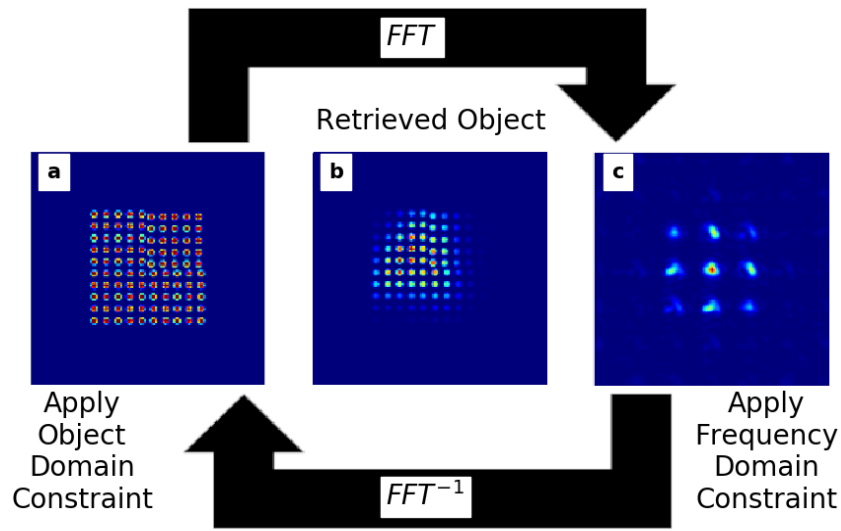

Fig. 1. **Iterative CDI Phase Retrieval**

### 3. Convolutional Neural Network

Our neural network differs from an object retrieved from a diffraction pattern using iterative CDI algorithm [9] because rather than retrieving the object which is the Fourier transform of the diffraction pattern, we retrieve the wavefront before it propagates through a mask (Fig. 2 c), our wavefront sensor. The neural network is able to perform an inference to retrieve the wavefront which is incident on the wavefront sensor (constraint). The neural network structure (Fig. 3) consists of convolutional and deconvolutional layers. The input to the neural network is the diffraction pattern. This input is down-sampled by several zero-padded convolutional layers with stride. The stride is the distance between points where the convolutional kernel is applied to the previous layer. Stride value greater than 1 is used to decrease the matrix size along the image (x,y) dimensions after passing through each convolutional layer. At the encoded layer (smallest x,y dimensions), there are two split branches of deconvolutional layers which up-sample the encoded layer to a real and imaginary output. Therefore, the diffraction pattern is input to the network, and the real and imaginary parts of the object which created the diffraction pattern are output. We use the real and imaginary part of the object rather than the amplitude and phase to avoid problems with phase wrapping and sharp discontinuities in the retrieved phase. We train the neural network with a dataset of simulated data. The data set for training a neural network using supervised learning consists of inputs and labels. In this case the input is the diffraction pattern. The output is what we want the neural network to retrieve from the input. Our labels are the corresponding complex objects (Fig. 2 c) which pass through the wavefront sensor (Fig. 2 d) to create the diffraction pattern that is input to the network. The training process for the neural network is the network learning a mapping of the input to the correct output. The training is done by minimizing a cost function (Eq. 1), which is the error between the output of the network and the corresponding label (real and imaginary object), and the error of the reconstructed diffraction pattern compared to the input diffraction pattern, flattened to 1-dimensional vectors.

$$C = \frac{1}{n}\sum_{i=1}^{n}(U_{\text{real}_i}^{\text{retrieved}} - U_{\text{real}_i}^{\text{actual}}) + \frac{1}{n}\sum_{i=1}^{n}(U_{\text{imag}_i}^{\text{retrieved}} - U_{\text{imag}_i}^{\text{actual}}) + \frac{1}{n}\sum_{i=1}^{n}(I_i^{\text{retrieved}} - I_i^{\text{actual}}) \quad (1)$$

$$I : \text{Diffraction Pattern}$$
$$U : \text{Object}$$

The use of a reconstructed diffraction pattern in the cost function has been shown to effectively improve the accuracy of neural network training related to coherent diffractive imaging [7]. The reconstructed diffraction pattern is calculated by simulating the propagation of the retrieved object through the wavefront sensor using the multi slice approach [13]. This training process is the computationally expensive part of using neural networks because the training process involves the calculation of many partial derivatives of the cost function with respect to all the weights of the artificial neurons in the network over many iterations. This process is highly parallelizable and is the reason the graphics card is highly effective for use in neural networks. Once the network is trained, using the network to produce an output from a diffraction pattern takes a very short time (milliseconds). Running on a Tesla V100 graphics card, the complex object is retrieved from a diffraction pattern in 23 milliseconds. In comparison to iterative phase retrieval, hundreds of iterations must be ran, which usually takes several seconds on a standard machine. The neural network is mathematically a series of matrix multiplications which can be ran in parralel on a GPU highly efficiently. Additionally, after the iterative retrieval has converged, the retrieved object must be interpolated to reveal the wavefront incident on the sensor, which takes additional computational time [10]. This interpolation is performed by the neural network in the single pass through the network.

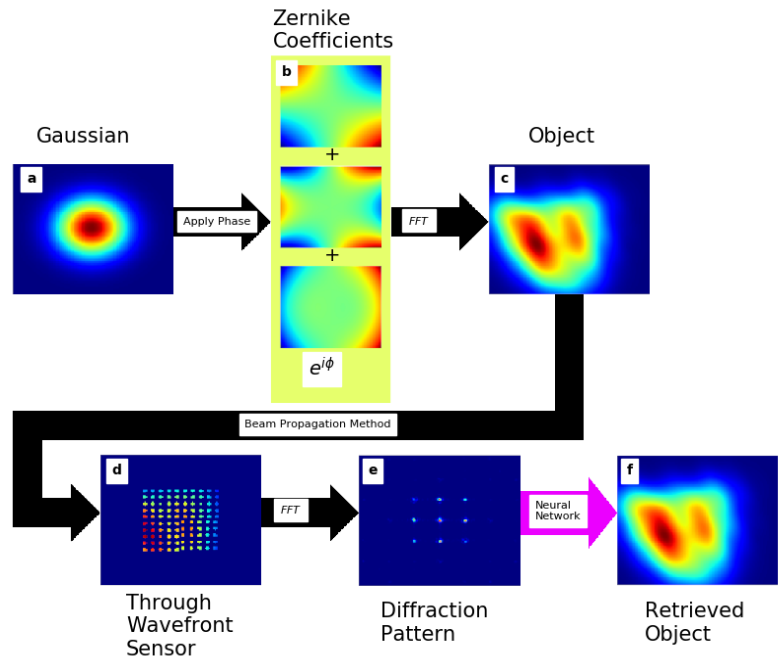

Fig. 2. **Object simulation and neural network** (a) Gaussian profile (b) Zernike polynomials which are applied to the Gaussian (c) Fourier transform of the Gaussian with applied phase (d) The output after propagation through the wavefront sensor (e) The fourier transform of the object after the wavefront sensor, and the diffraction pattern which is imaged by the detecter (f) The object retrieved by the neural network

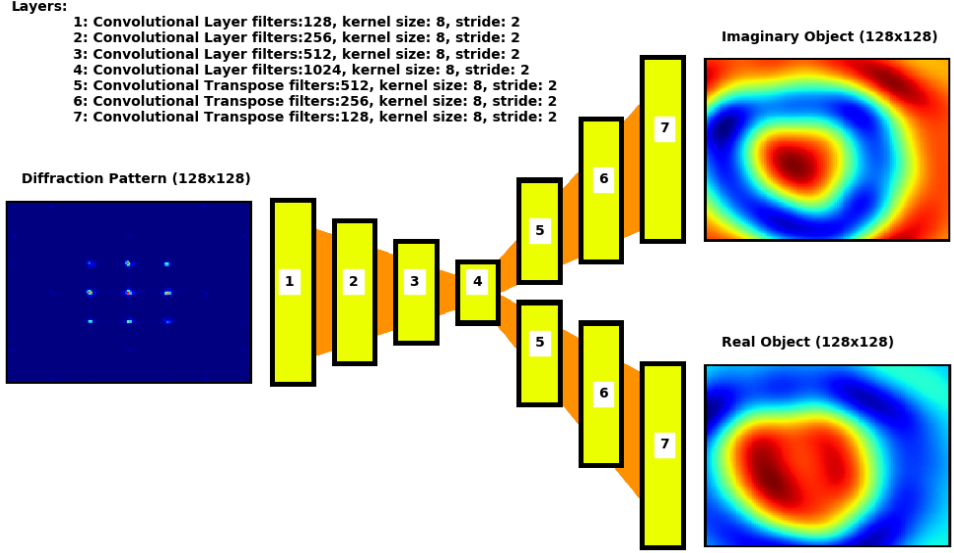

Fig. 3. **Neural Network Structure** The diffraction pattern is input to the encoder, which consists of 4 convolutional layers with a stride of two, and increasing number of output channels used to down-sample the input. The signal is up-sampled by two series of convolutional transpose operations which output the real and imaginary part of the retrieved object

We generate our training data set by simulating aberrations described by Zernike polynomials. We apply randomized aberrations to a beam and calculate the the shape of the beam in the focus by applying a Fourier transform. The randomized aberrations are calculated by multiplying the Zernike polynomials by a random scalar number to create a phase (Eq. 2), and applying it to a gaussian amplitude.

$$\phi_{\text{zernike}}(x,y) = C_1 \cdot z_2^2(x,y) + C_2 \cdot z_2^0(x,y) + C_3 \cdot z_2^2(x,y) + \ldots + C_{11} \cdot z_4^2(x,y) + C_{12} \cdot z_4^4(x,y) \quad (2)$$

$z_n^m$ : zernike polynomial of m and n order
$C_i$ : randomized scalar for zernike polynomial

The Gaussian with applied phase is propagated to the far-field with Fourier transform (Eq. 3).

$$\hat{E}_{\text{wavefront}}(x,y) = FFT\left[\exp\left(\frac{-x^2 - y^2}{w_0^2}\right) \cdot \exp\left(i \cdot \phi_{\text{zernike}}\right)\right] \quad (3)$$

This object is then propagated through the wavefront sensor $\hat{M}$ using the multi slice BPM technique [13] (Eq. 4).

$$\gamma(f_x, f_y) : \sqrt{1 - (\lambda \cdot f_x)^2 - (\lambda \cdot f_y)^2}$$

$$\hat{E}^{i+1}(x, y) = FFT^{-1}\left[FFT\left[\hat{E}^{i}(x, y) \cdot \hat{M}_{\text{wavefront sensor}}(x, y)\right] \cdot \exp\left(i\frac{2\pi \cdot dz}{\lambda} \cdot \gamma(f_x, f_y)\right)\right] \quad (4)$$

The propagation takes into account the wavelength of the laser beam, which also shapes the produced diffraction pattern. The diffraction pattern is produced by Fourier transforming the propagated beam. The label is the complex object (Fig. 2 c), and the input is the produced diffraction pattern (Fig. 2 e). By constructing the training data set in this way, we build in the constraint to the neural network. Because we propagate all the objects through the same wavefront sensor, the neural network learns to retrieve the objects behind this specific wavefront sensor. This is analogous to using a fixed support in the iterative CDI retrieval method.

It is possible for different objects to produce an identical diffraction pattern (ambiguity) when the object contains a constant phase shift. An ambiguity is contained in a data set when there are multiple diffraction patterns which are (nearly) identical, and correspond to different wavefronts. Ambiguities in a neural network data set must be removed to train the neural network, or the training will not be able to converge on the solution. We remove ambiguities in the wavefront phase by setting the phase angle of the complex object to 0 at the center for each wavefront in the training data set. In this way, the neural network learns the convention of setting the phase angle to 0 at the center of the retrieaved wavefront, and data set does not consist of ambiguities due to constant phase shift. The twin image problem [14], which also imposes an ambiguity, is solved by choosing an asymmetrical geometry of the binary mask (wavefront sensor).

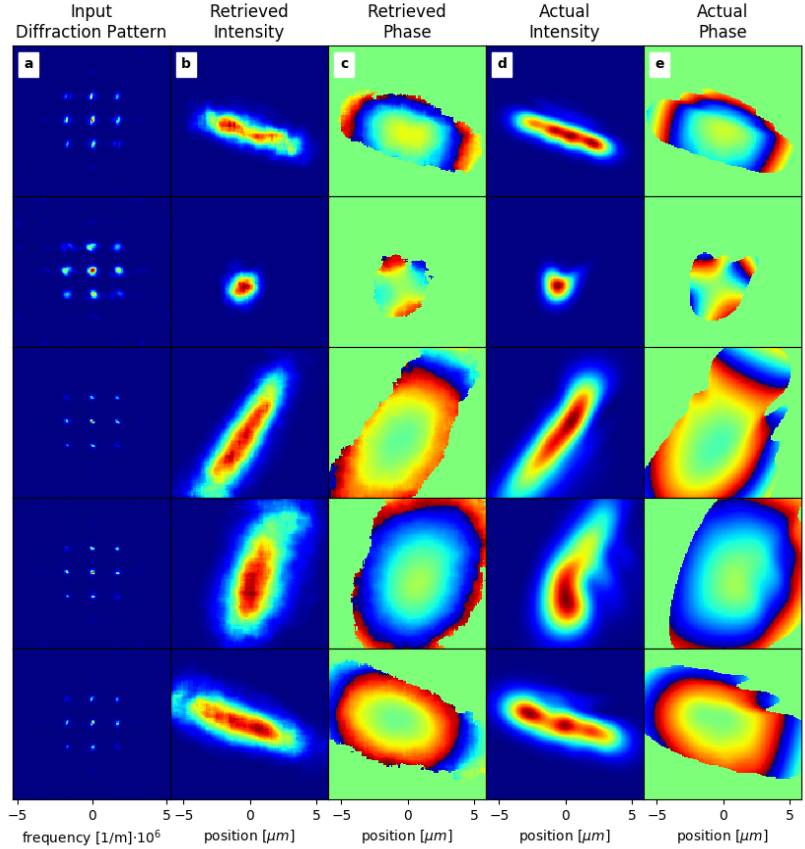

Fig. 4. **Neural Network Retrieval on Validation data** (a) Input Diffraction Pattern (b) Retrieved Intensity (c) Retrieved Phase (d) Actual Intensity (e) Actual Phase

Two data sets are constructed using the model shown in Figure 2, a training data set and an validation data set. The training data set consists of 36000 samples, and the validation data set consists of 200 samples. We use the training data set to calculate gradients of the cost function with respect to all the weights in the neural network, and adjust the weights to minimize the cost function using ADAM optimization [15]. The validation set is used to determine when to stop training the network. We stop the training when the error on the validation data starts to increase, while the error on the training data continues to decrease. This determines whether the network is over-fitted to the data set. The validation data set is representative of actual measured diffraction patterns because the network is not trained with this data.

## 4. Experimental Results

We test our trained neural network on experimentally measured data. An XUV source is used to illuminate a pinhole of diameter 2.7 $\mu m$. We use the neural network to retrieve the beam which is propagated through the pinhole. The wavefront sensor is placed behind the pinhole (Fig. 5) and we measure the diffraction pattern produced by the wavefront sensor placed various distances from the pinhole. The detector is placed behind the wavefront sensor at a fixed distance such that the measured diffraction pattern is sufficiently measured in the far field, and Fraunhofer Diffraction is observed.

The diameter of the retrieved beam increases as the distance between the wavefront sensor and the pinhole is increased. We verify that our retrieved object is correct by using the near field propagator (Eq. 5).

$$E_{\text{prop}}(x, y) = FFT^{-1}\left[FFT\left[E(x, y)\right]e^{i\frac{2\pi z}{\lambda}\gamma(f_x, f_y)}\right] \tag{5}$$

We know the diameter of the pinhole is 2.7 $\mu m$ from an electron microscope image, so we are able to verify the accuracy of our retrieval by propagating the retrieved back to the pinhole. We propagate the retrieved complex object which is retrieved at a distance of 500 $\mu m$ from the pinhole (Fig. 6 (h,i)). The retrieved wavefront is propagated by a distance of 500 $\mu m$ toward the spherical aperture (Fig. 6 (j,k)), and we find that it matches the 2.7 micrometer diameter of our pinhole (Fig. 6).

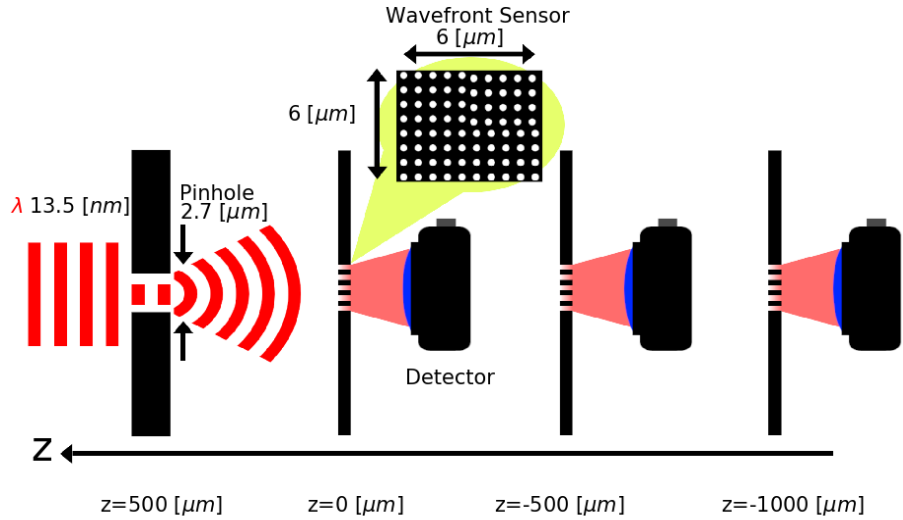

Fig. 5. **Experimental setup** The Detector is placed behind a pinhole of diameter 2.7 $\mu m$ with an incoming XUV beam of wavelength 13.5 $nm$. The distance between the pinhole and the wavefront sensor is adjusted.

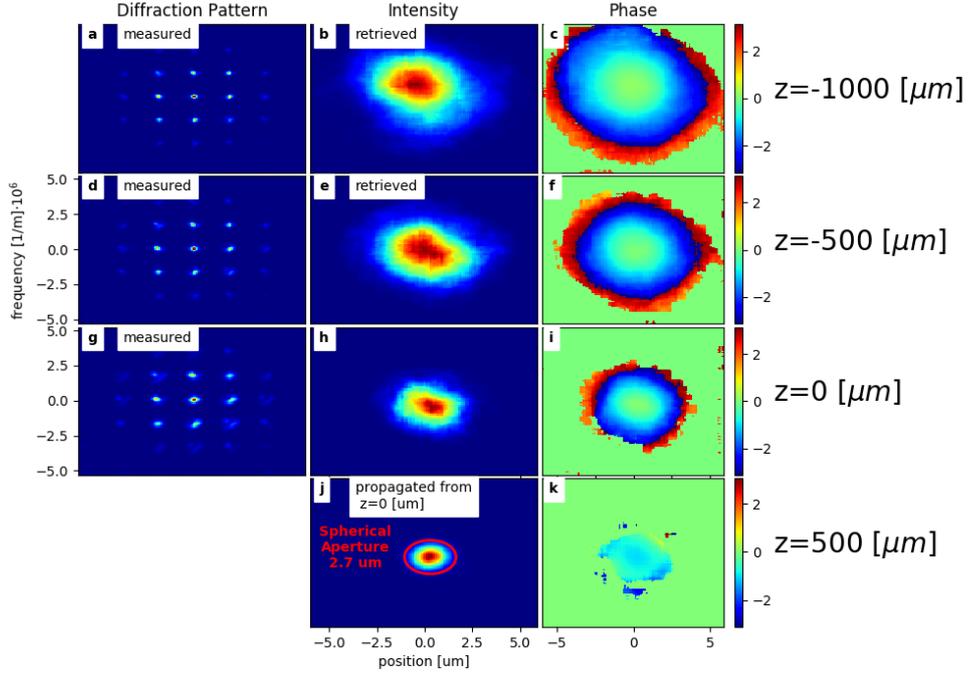

Fig. 6. **Retrieval with Neural Network at various distances behind pinhole** The trained neural network is used to retrieve the object behind the pinhole. The retrieved object at 0 *μm* meters is propagated a distance of 500 *μm* to the pinhole.

The neural network is able to retrieve the object without an iterative retrieval, and it retrieves the object behind the wavefront sensor, so there is no need to perform interpolation of the object to get the wavefront, as is the case with iterative CDI retrieval.

## 5. Noisy Diffraction Patterns

We test the neural network and the iterative phase retrieval method on simulated samples with noise applied to simulate a camera image captured from a weak laser source with limited number of photons. The finite number of counts measured by the detector is characterized mathematically by the Poisson distribution. The neural network is trained on a training data constructed from 36000 unique objects. For each object the neural network is trained on the corresponding diffraction pattern with no noise (infinite counts), 50, 40, 30, 20, 10, and 5 peak signal counts. 200 samples which were not used during the neural network training are retrieved using both the neural network and the iterative phase retrieval method at various signal to noise (SNR) levels. The RMSE error of these retrievals is calculated by measuring the difference between the actual intensity/phase and the retrieved intensity/phase. The neural network outperforms the iterative phase retrieval as shown in Figure 7.

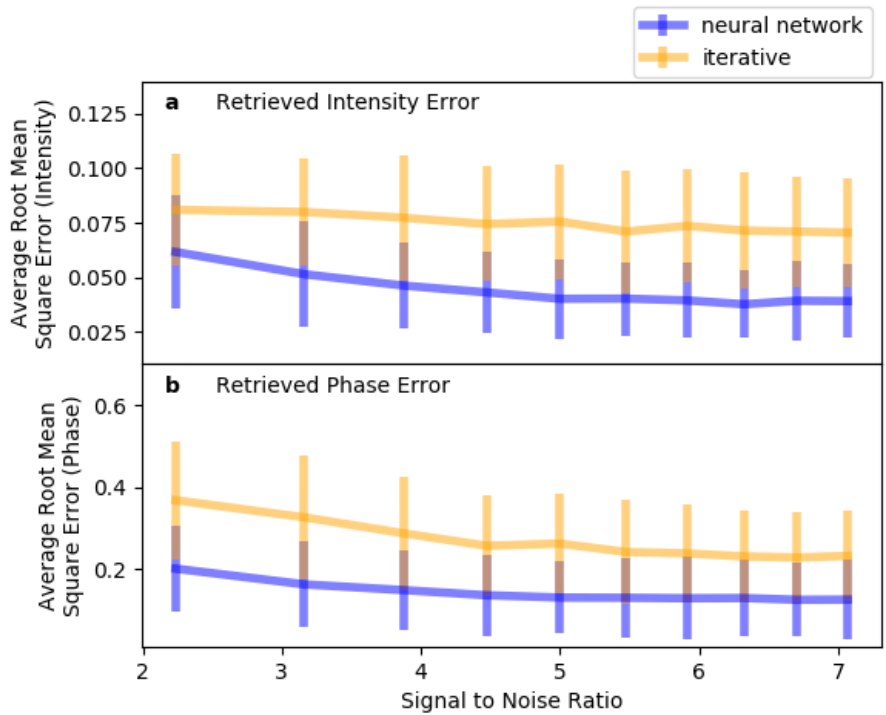

Fig. 7. **Retrieval on noisy diffraction patterns compared** Both the iterative and neural network methods are used to retrieve intensity (a) and Phase (b) from diffraction patterns with an increasing amount of noise characterized by the SNR of the measured diffraction pattern. The Average retreival error is plotted along with the standard deviation.

The retrieved object from both the neural network and the iterative phase retrieval method is shown in Figure 8. In the high noise case with the diffraction pattern imaged with 10 peak counts (3.16 SNR), the neural network is still able to retrive the object with high accuracy (Fig. 8 i,j), and the object retrieved by the iterative method (Fig. 8 k,l) does not resemble the actual object.

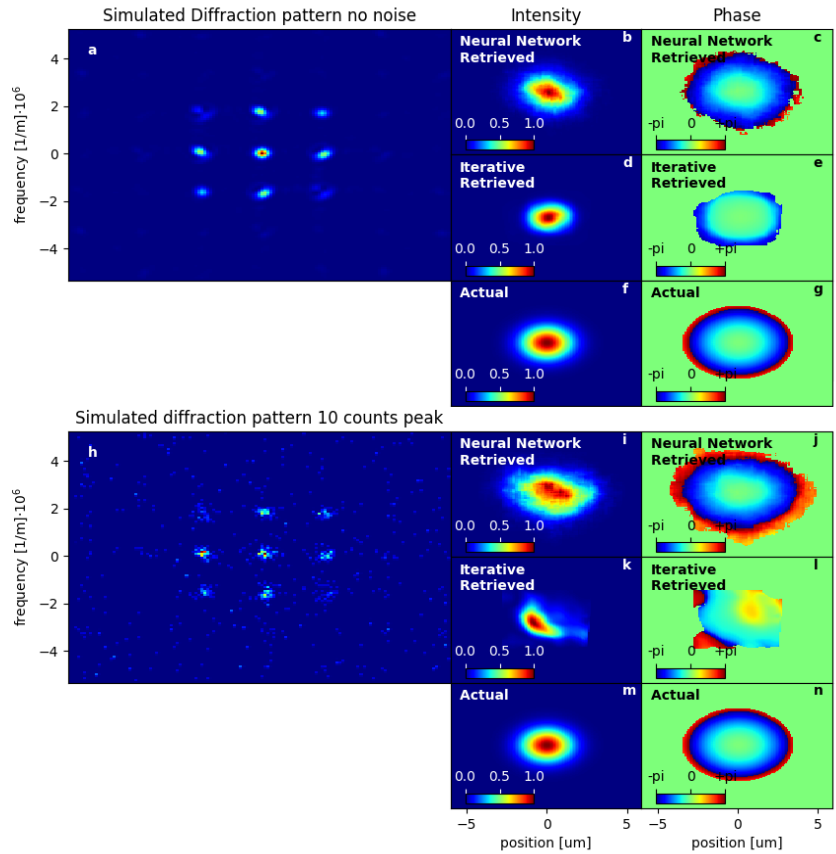

Fig. 8. **Diffraction pattern with and without noise retrieved** An identical object (f,g,m,n) is used to construct a diffraction pattern with (h) and without (a) noise. The object is retrieved from the noise free diffraction pattern (a), the result of the neural network (b,c) and the iterative retrieval (d,e) are shown, and compared to the actual object (f,g). The object is retrieved from the noisy diffraction pattern (h), the results of the neural network (i,j) and the iterative retrieval (k,l) are compared to the actual object (m,n)

The accuracy of the neural network retrieval in high noise diffraction patterns in combination with the speed of retrieval makes the neural network method highly advantageous when working in an environment with low intensity beams, the integration time of capturing the image can be shorter, even enabling single shot measurements with a small fraction of the beam.

## 6. Conclusion

In conclusion, we demonstrate a new approach to coherent diffractive imaging which works in conjunction with wavefront sensing using a mask. The retrieval time has been reduced from seconds to milliseconds, which will enable live in-focal wavefront measurements, even in very weak beams which produce a noisy diffraction pattern with a short integration time.

The wavefront sensor can be applied to a broad range of wavelengths ranging from visible light to X-ray. Potentially this method can be applied to beam characterization in high harmonic generation, Synchrotrons and free electron lasers. Additionally, the mask wavefront sensor is a cost effective alternative to commercial wavefront sensors which are more difficult to align. Using the mask wavefront sensor, only the camera-sample distance must be known.

**Funding**

Federal State of Thuringia (2017 FGR 0076); European Social Fund (ESF) Financial support by the Thuringian State Government within its ProExcellence initiative (APC2020) is acknowledged. The authors declare no conflicts of interest

**References**

1. S. Lawrence, C. Giles, A. Tsoi, and A. Back, "Face recognition: A convolutional neural network approach," Neural Networks, IEEE Transactions on **8**, 98 – 113 (1997).
2. B. Djavan, M. Remzi, A. Zlotta, C. Seitz, P. Snow, and M. Marberger, "Novel artificial neural network for early detection of prostate cancer," J. Clin. Oncol. **20**, 921–929 (2002). PMID: 11844812.
3. L. Bottaci, P. J. Drew, J. E. Hartley, M. B. Hadfield, R. Farouk, P. W. Lee, I. M. Macintyre, G. S. Duthie, and J. R. Monson, "Artificial neural networks applied to outcome prediction for colorectal cancer patients in separate institutions," The Lancet **350**, 469 – 472 (1997).
4. Z. Zhu, J. White, Z. Chang, and S. Pang, "Attosecond pulse retrieval from noisy streaking traces with conditional variational generative network," Sci. Reports **10** (2020).
5. T. Zahavy, A. Dikopoltsev, O. Cohen, S. Mannor, and M. Segev, "Deep learning reconstruction of ultrashort pulses," (2018), p. STh4N.1.
6. M. Cherukara, Y. Nashed, and R. Harder, "Real-time coherent diffraction inversion using deep generative networks," Sci. Reports **8** (2018).
7. Z. Guan, E. H. Tsai, X. Huang, K. Yager, and H. Qin, "Ptychonet: Fast and high quality phase retrieval for ptychography," in *BMVC,* (2019).
8. Y. Nishizaki, R. Horisaki, K. Kitaguchi, M. Saito, and J. Tanida, "Analysis of non-iterative phase retrieval based on machine learning," Opt. Rev. (2020).
9. J. R. Fienup, "Phase retrieval algorithms: a comparison," Appl. optics **21**, 2758–2769 (1982).
10. W. Eschen, G. Tadesse, Y. Peng, M. Steinert, T. Pertsch, J. Limpert, and J. Rothhardt, "Single-shot characterization of strongly focused coherent xuv and soft x-ray beams," Opt. Lett. **45**, 4798–4801 (2020).
11. M. P. Benk, K. A. Goldberg, A. Wojdyla, C. N. Anderson, F. Salmassi, P. P. Naulleau, and M. Kocsis, "Demonstration of 22-nm half pitch resolution on the sharp euv microscope," J. Vac. Sci. & Technol. B, Nanotechnol. Microelectron. Materials, Process. Meas. Phenom. **33**, 06FE01 (2015).
12. W. Chao, B. D. Harteneck, J. A. Liddle, E. H. Anderson, and D. T. Attwood, "Soft x-ray microscopy at a spatial resolution better than 15 nm," Nature **435**, 1210–1213 (2005).
13. G. Tadesse, W. Eschen, R. Klas, M. Tschernajew, F. Tuitje, M. Steinert, M. Zilk, V. Schuster, M. Zuerch, T. Pertsch, C. Spielmann, J. Limpert, and J. Rothhardt, "Wavelength-scale ptychographic coherent diffractive imaging using a high-order harmonic source," Sci. Reports **9** (2019).
14. M. Guizar-Sicairos and J. R. Fienup, "Understanding the twin-image problem in phase retrieval," J. Opt. Soc. Am. A **29**, 2367–2375 (2012).
15. D. Kingma and J. Ba, "Adam: A method for stochastic optimization," Int. Conf. on Learn. Represent. (2014).